\documentstyle[preprint,pre,aps]{revtex}
\begin{document}
\preprint{TNT 95-ShPre-V.9}
\draft
\title{Predictability in Systems with Many Characteristic Times:
       The Case of Turbulence}

\author{E. Aurell}
\address{Department of Mathematics, Stockholm University
          S--106 91 Stockholm, Sweden}

\author{G. Boffetta}
\address{Istituto di Fisica Generale, Universit\`a di Torino,
         Via Pietro Giuria 1, I-10125 Torino, Italy}

\author{A. Crisanti}
\address{Dipartimento di Fisica, Universit\`a di Roma ``La Sapienza'',
         P.le Aldo Moro 2, I-00185 Roma, Italy}

\author{G. Paladin}
\address{Dipartimento di Fisica, Universit\`a dell'Aquila
         Via Vetoio, Coppito I-67100 L'Aquila, Italy}

\author{A. Vulpiani}
\address{Dipartimento di Fisica, Universit\`a di Roma ``La Sapienza'',
         P.le Aldo Moro 2, I-00185 Roma, Italy}

\date{\today}

\maketitle

\begin{abstract}
In chaotic dynamical systems, an infinitesimal
perturbation is exponentially amplified at a time-rate 
 given by the inverse of the maximum  Lyapunov exponent $\lambda$. 
In fully developed turbulence, $\lambda$ grows
as a power of the Reynolds number.
This result could seem in contrast with phenomenological
 arguments suggesting that,
  as a consequence of `physical'  perturbations, 
the predictability time is  roughly given by 
 the characteristic life-time of the large  scale structures,
and hence independent of the Reynolds number.
We show that such a situation is present in 
  generic systems with many
degrees of freedom, since  the growth of a non-infinitesimal 
 perturbation 
 is determined by cumulative effects of many different
characteristic times and is unrelated to the maximum Lyapunov exponent. 
Our results are illustrated in a chain of coupled maps
and in a shell model for the energy cascade in turbulence.
\end{abstract}

%\pacs{05.45.+b - Theory and models of chaotic systems, 
%      47.27.Gs Isotropic turbulence; homogeneous turbulence}

\section{Introduction}
\label{sec:Intro}
After the seminal work of Lorenz \cite{Lorenz63}, it is well understood that 
 the predictability of the state of a system ruled by a deterministic evolution 
law  has severe limitations in presence of  deterministic chaos.
In systems with  sensitive dependence on the initial condition one has
an exponential divergence of the distance $\delta {\bf x}$ between two initially
close trajectories, i.e.
\begin{equation}
|\delta {\bf x}(t)|\simeq |\delta {\bf x}(0)|\, e^{\lambda t}
\label{eq:dx}
\end{equation}
where $\lambda$ is the maximum Lyapunov exponent \cite{Benettin}.
Consequently, if $|\delta {\bf x}(0)|=\delta_0$ and one accepts 
a maximum tolerance $\delta_{max}$ on the knowledge of the state of the system,
(\ref{eq:dx}) implies that the systems is predictable up to a time
\begin{equation}
T \sim {1\over \lambda}\, \ln\left({\delta_{max} \over \delta_0}\right).
\label{eq:tlyap}
\end{equation}

Equation (\ref{eq:tlyap}) gives  only a first rough answer to the problem 
since it does not take into account some important features of
chaotic systems.
Endeed to study the predictability of a generic dynamical system
one has to consider the following non trivial aspects:

\begin{itemize}

\item{a)} The Lyapunov exponent $\lambda$ is a global quantity: 
it measures the average exponential rate of divergence of nearby 
trajectories. In general there are finite-time fluctuations of this 
rate and it is possible to define an `instantaneous' rate $\gamma$, called 
effective Lyapunov exponent \cite{PV87}, which depends 
on the particular point of the trajectory
${\bf x}(t)$ where the perturbation is performed. In the same way, 
the predictability time $T$  fluctuates, following the $\gamma$-variations.

\item{b)} In dynamical systems with many degrees of freedom, the interactions
among different parts of the system play an important role on the 
growth of a perturbation. The statistics of the effective 
Lyapunov exponent is not sufficient to characterize
 the growth of infinitesimal perturbations 
 and one has to analyze the behavior of the tangent 
vector ${\bf z}(t)$, i.e. of the direction along which an infinitesimal 
perturbation grows, see e.g \cite{Pikovsky93}.
Moreover, if  one is interested in the behaviour of a perturbation
concentrated on certain degrees of freedom, e.g. small length
scales in weather forecasting, and in a prediction on the evolution
of other degrees of freedom, e.g. large length scales,
a relevant quantity is the time $T_R$ necessary to the tangent vector
to relax on the time dependent eigenvector $\bbox{e}(t)$ 
of the stability matrix, corresponding to the maximum Lyapunov exponent.
If the  perturbations  are small enough, i.e. $\delta {\bf x}$ is 
proportional to the tangent vector, one has that
\begin{equation}
T \sim T_R \, + \, {1\over \lambda} \, \ln {\delta_{max}\over \delta_0} 
\label{eq:tr}
\end{equation}
where in general $T_R$ may depend on $\delta{\bf x}$. So the mechanism 
of transfer of the error $\delta{\bf x}$ through the degrees of freedom 
of the system could be more important than the rate of divergence of 
nearby trajectories \cite{PalVul}.

\item{c)} In systems with many characteristic times, such as the eddy 
turn-over times in fully developed turbulence, if the perturbation is 
not infinitesimal, or if the threshold of accepted error is not
small, $T$ is determined by the detailed mechanism of transfer due to the  
non-linear effects in the evolution equation for $\delta{\bf x}$.
In this case, the predictability time might have no relation with the 
Lyapunov exponent and $T$ depends in a non-trivial way on the details 
of the system.

\end{itemize}

The aspects a) and b) have been studied in previous works
\cite{PalVul,CJPV93,PRL}. In this paper 
we mainly address point c). We investigate the
predictability problem for non-infinitesimal perturbations in
the framework of two
models: a system of coupled maps, and a shell model for the energy cascade 
in three dimensional turbulence. 

The paper is organized as follows.
In section \ref{sec:Somer} we discuss some phenomenological results in 
fully developed turbulence. In section \ref{sec:Maps} the
predictability problem is discussed in a system of coupled maps,
 using analytic methods supported by  some numerical investigations. 
Section \ref{sec:Shellm} analyzes the more realistic case of
 a shell model of turbulence by both  
numerical simulations and closure approximations.
In section \ref{sec:Discussion} we discuss the physical relevance of our 
results and open problems.

\section{Some results for the predictability in fully developed turbulence}
\label{sec:Somer}
In three dimensional fully developed turbulence, the inverse maximum 
Lyapunov exponent is roughly proportional to the smallest 
characteristic time of the system, the turn-over time $\tau$ of eddies 
of the size of the Kolmogorov length $\eta$. The argument, due to Ruelle 
\cite{Ruelle79}, is the following. The Kolmogorov theory predicts
that the longitudinal velocity difference at distance $\ell=|{\bf r}|$
scales as
\begin{equation}
v(\ell) \equiv |{\bf v}({\bf x} + {\bf r})-{\bf {v}}({\bf x})| 
\sim \epsilon^{1/3}
\ell^{2/3} \sim V \, \, \left({\ell \over L}\right)^{1/3}
\label{eq:dvl}
\end{equation}
where $V$ and $L$ are the typical velocity and length of the energy
containing eddies, and $\epsilon$ is the mean  rate of 
 energy dissipation.

The non-linear transfer of energy is stopped at the Kolmogorov scale $\eta$
where viscosity $\nu$ is able to compete with the convective term, i.e.
$\nu\sim \eta\,v(\eta)$, thus from (\ref{eq:dvl}) we have
\begin{equation}
\eta \sim L \,  Re^{-3/4}
\label{eq:eta}
\end{equation}
where $Re = V L /\nu$ is the Reynolds number.
The corresponding turn-over time is
\begin{equation}
\tau(\eta)\sim \frac{\eta}{v(\eta)}
          \sim T_0 \left({ \eta \over L}\right) ^{2/3}
\label{eq:taueta}
\end{equation}
where $T_0=L/V$ is the life time of the large scale disturbances.

These dimensional relations imply that the maximum Lyapunov exponent 
scales with $Re$ as
\begin{equation}
\lambda \sim {1 \over \tau(\eta)}
             \sim {1\over T_0}\,Re^ {1/2}
\label{eq:lambda}
\end{equation}

Taking into account the intermittency one expects that 
the presence of quiescent quasi-laminar periods
changes the chaotic features of the fluid flow.
The intermittency of energy dissipation can be described by introducing
a spectrum of singularities $h$ of the velocity gradients \cite{ParFri},
 i.e. assuming a local scaling invariance
 so that in any point ${\bf x}$ of the fluid 
 $v(\ell) \sim \ell^{h({\bf x})}$.
 In the framework of the
multifractal approach, one thus finds \cite{PRL}
\begin{equation}
\lambda \sim {1\over T_0}\,Re^\alpha
       \qquad \hbox{\rm with}\
     \alpha=\max_h \,\left[\, {D(h)-2-h \over 1+h}\,\right]
\label{eq:lambdamf}
\end{equation}
where $D(h)$ is the fractal dimension of the set of fluid points
characterized by a given singularity $h$. 
The value of $\alpha$ depends on $D(h)$. By using
the function $D(h)$ obtained by fitting the exponents $\zeta_q$ 
of the velocity structure functions
with the random beta model \cite{BPPV84}, one has $\alpha=0.459...$, 
slightly smaller than the Ruelle prediction $\alpha=0.5$, see (\ref{eq:lambda})

Relation (\ref{eq:tlyap}) and (\ref{eq:lambdamf}) tell us that considering 
a very small perturbation
at $t=0$ and a very small tolerance $\delta_{max}$, the predictability time 
vanishes as the Reynolds number increases, as verified
in recent numerical simulations 
of the shell model (see sect. \ref{sec:Shellm}) \cite{CJPV93}.

On a very different ground, without considering the 
 exponential growth of infinitesimal disturbances,
Lorenz \cite{Lorenz69}, see also \cite{Lilly73} and \cite{LK72},
proposed a phenomenological approach to the predictability
problem in turbulence. Consider
wave-numbers around $k$ with corresponding typical spatial scale $\ell
\sim k^{-1}$. The time $\tau(k)$ for a perturbation at wavenumber $2 k$ 
to induce a complete uncertainty on the velocity field on the wave 
number $k$ is assumed to be proportional to the typical eddy turn-over 
time at scale $\ell$, from (\ref{eq:dvl}):
\begin{equation}
\tau(\ell) \sim { \ell \over v(\ell)} \sim T_0 \, \left({\ell \over
L}\right)^{2/3}.
\label{eq:taulorenz}
\end{equation}

An incertitude $O(v(\eta))$ propagates through an inverse cascade
from the Kolmogorov scale $\eta \sim k_d^{-1}$ up to the scale of the energy
containing eddies $L \sim k_0^{-1}$. The predictabilty time on the 
large scales is therefore 
\begin{equation}
T = \sum_{n=0}^{N} \tau(2^{n} \eta)
\label{eq:tsum}
\end{equation}
where $N=\log_{2}(k_d/k_0) \sim \log Re$. The geometric series
(\ref{eq:tsum}) is dominated by the term $n=N$ so that $T$ is practically 
independent of Reynolds number:
\begin{equation}
T \sim T_0 \sim L/V.
\label{eq:tsum2}
\end{equation}

It is worth remarking that closure approximations \cite{Lilly73,LK72} 
allows one to write down simple equations for the evolution
 of disturbances in turbulent flows and fully confirm these results.
In appendix B, we derive this type of equations in the simplified
 framework of the shell model, but there are no conceptual differences
 with the analogous derivation in 
 the case of the Navier-Stokes equations.

The Lorenz picture of an inverse cascade of a perturbation
 also permits to estimate the growth of a perturbation at intermediate times.
Indeed, after a time $t$, a perturbation localized on the Kolmogorov length 
scale, is expected to affect an eddy of size 
$\ell_t$ such that its turn-over time is $\tau(\ell_t) \sim t$.
It follows \cite{Lesieur}, see (\ref{eq:taulorenz})
\begin{equation}
\ell_t \sim L \, \left({t \over T_0}\right)^{3/2}
\label{eq:scalet}
\end{equation}
Let us indicate by $\bbox{v}$ the reference velocity field and
by $\bbox{v}'$ the perturbed field.
The size of the difference  $\delta \bbox{v}(t)$
between two velocity fields at time $t$ 
can be estimated as the typical velocity of a disturbance at scale $\ell_t$,
so that from (\ref{eq:dvl}) and (\ref{eq:scalet})
\begin{equation}
|\delta \bbox{v}(t)| \sim v(\ell_t) \sim V \, \sqrt{t/T_0}.
\label{eq:dvt}
\end{equation}
Perturbations grow as square root of
time. 
Let us stress again that such perturbations cannot be
considered infinitesimal so that the inverse cascade picture
is not in contrast
 with the exponential amplification of errors on initial
 conditions that is present in  chaotic dynamical systems.

There are two main predictions of the Lorenz approach:
predictability time on large scale independent of Reynolds number 
and growth of perturbations as a power of time instead
of an exponential. 

In the next sections we use the theory of dynamical systems 
with many degrees of freedom to argue 
that while the former is correct,   the latter is 
a too rough description of the real behavior
of $\delta \bbox{v}(t)$.

\section{Predictability in a chain of coupled maps}
\label{sec:Maps}

In this section we discuss the predictability problem in a system
of coupled maps each with a different time scale. 
This is an idealized situation where each degree of freedom has
a chaotic dynamics with its own characteristic time and it is coupled to
the other degrees of freedom by a weak local interaction.
Our toy model can be considered the  prototype for physical 
situations where one can separate the evolution on different scales.
Despite its simplicity, it displays non-trivial
properties which enlighten the behaviour of more complex systems.

The system is given by a chain of chaotic coupled maps
\begin{equation}
\label{eq:guno}
\left\{
\begin{array}{ll}
x_1(t+1) &=  (1-\epsilon) g_1(x_1(t)) + \epsilon f(x_2(t)) \nonumber \\
x_2(t+1) &=  (1-\epsilon) g_2(x_2(t)) + \epsilon f(x_3(t)) \nonumber \\
           &\cdots   \\
x_{N-1}(t+1) &=  (1-\epsilon) g_{N-1}(x_{N-1}(t)) + 
\epsilon f(x_N(t)) \nonumber \\
x_{N}(t+1) &=  g_N(x_{N}(t)) \nonumber 
\end{array}\right.
\end{equation}
where the $N$ variables ${x_k}$ are defined on the interval $I=[0,1]$,
  $\epsilon$ is the coupling constant, $g_k$ and $f$ are
functions of $I$ into itself. In order to mimic the energy cascade
 of turbulent flow, 
we assume that the function $g_k$ represents  
the evolution law of an eddy of size $l_k$
in the absence of interactions. In this context, 
$x_k$ can be regarded as the velocity difference of the eddies on the 
octave of length scale $l_k=L_0 2^{-k}$. 
The evolution 
is chaotic with a Lyapunov exponent proportional to the inverse
of the eddy turn-over time $\tau_k$. 
The simplest possible choice for $g_k(x)$ is the piecewise linear map 
\begin{equation}
g_k(x) = e^{\lambda_k} x \qquad {\rm mod}\ 1
\label{eq:guno3}
\end{equation}

From dimensional arguments, and using
(\ref{eq:taulorenz}) we can estimate the eddy turn-over time
 $\tau_k$ so that the Lyapunov exponent is 
 \begin{equation}
\lambda_k = {1 \over T_0} \, 2^{2/3(k-1)}
\label{eq:gdue}
\end{equation}

The coupling parameter $\epsilon$ is small and constant for each
scale, although more realistic models could consider
scale-dependent couplings. The form of $f(x)$ --
 the term representing the local interaction between eddies -- 
 does not affect the qualitative results. 
We have considered two possibilities:  $f(x)=x$ and $f(x)=x^2$. 

In order to study the predictability, 
the system (\ref{eq:guno}) has been integrated with
two different initial conditions ${\bf x}(0)$ and ${\bf x}'(0)$ 
differing on the smallest scale, 
$\Delta_k(0) \equiv x'_k(0)-x_k(0) =\delta_{k,N} \delta_0$. The initial
incertitude is thus confined to the smallest and fastest scale $x_N$ while
we want to forecast the behavior of the system at the largest scale $x_1$. 

Because time-scales are well separated, we expect that at the
beginning the uncertainty in the system is driven by the fastest scale
$x_N$ and the error at any scale grows with the smallest characteristic
time 
\begin{equation}
\Delta_k(t) \sim  \epsilon^{N-k} e^{\lambda_N t} \delta_0
\label{eq:gtre}
\end{equation}
where the power of $\epsilon$ is due to the locality of the
interactions. The behavior (\ref{eq:gtre}) will last up to the time $T_N$
at which the perturbation at small scale saturates, i.e. 
$\Delta_N = \Delta_{sat}$.
Then a second regime drived by the variable 
$x_{N-1}$, which has now the fastest exponential growth of errors, sets in. 
 When this second regime holds, 
 the incertitudes at scales $k \leq N-1$ grows as 
$\sim e^{\lambda_{N-1} t}$ until, at time $T_{N-1}$,
the variable $x_{N-1}$ saturates and $x_{N-2}$ dominates, and so on.
 Such an argument suggests that a perturbation 
$\Delta_1$ at the largest scale $x_1$ follows different exponential 
laws with rates $\lambda_k$ ($k=1,\cdots,n$)
 in each different regimes, with a global (envelope) evolution which 
 appear very similar to a power law.

Figure \ref{fig:fig1} shows the error observed at large scale $\Delta_1$ 
as a function of time in a numerical experiment realized with a chain of $N=10$
maps. The coupling function is chosen linear, $f(x)=x$, the coupling
parameter is $\epsilon=10^{-4}$, the largest characteristic time is
$T_0 = 200$. The initial conditions are random and the
initial perturbation on the small scale is $\delta_0=10^{-8}$. 
Figure \ref{fig:fig1} shows the error observed at large scale $\Delta_1$ 
as a function of time. Some exponential regimes are still
recognizable, the straight lines with different slopes in the
linear-logarithmic plot.

The observed behaviour can be well understood by means of a quasi-linear
analysis which also allows for some analytical estimates.
The evolution with (\ref{eq:guno3}) and linear $f(x)$ 
leads to a linear evolution for the perturbation,
\begin{equation}
{\bf \Delta}(t+1) = A \, {\bf \Delta}(t)
\label{eq:gquattro}
\end{equation}
where $A$ is the $N\times N$ Jacobian matrix 
\begin{equation}
A = \left( \begin{array}{cccccc}
l_1 & \epsilon & 0 & \ldots & 0 & 0 \\
0 & l_2 & \epsilon & \ldots & 0 & 0 \\
\vdots & \vdots & \vdots & \ddots & \vdots & \vdots \\
0 & 0 & 0 & \ldots & 0 & l_N
\end{array}
\right).
\label{eq:gcinque}
\end{equation}
Let us denote with $T_k$ the time at which scale $k$ saturates, i.e.
$\Delta_k(T_k)=1$, and with $M_k$ the period (number of steps) during
which scale $k$ dominates the dynamics, i.e. $M_k=T_{k}-T_{k+1}$.
If we suppose that the main contribution to the growth of perturbations
during period $M_k$ is given by the faster scale $\Delta_k$, which is
correct whenever $\epsilon$ is very small, we can write for any $i<k$
\begin{equation}
\Delta_i(T_k) = \Delta_i(T_{k+1}) + (A^{M_k})_{i,k} \Delta_k(T_{k+1}) 
\label{eq:gsei}
\end{equation}
while $\Delta_i(T_k)=\Delta_{sat}$ for $i \ge k$. 
The matrix $A^M$ can be easily computed and at the leading order is given by
\begin{equation}
(A^{M})_{i,k} = \epsilon^{k-i} {e^{M \lambda_{k}} - e^{M \lambda_{i}}
\over e^{\lambda_{k}} - e^{\lambda_{i}}}
\label{eq:gsette}
\end{equation}
The saturation periods are given by 
\begin{equation}
M_k = {1 \over \lambda_k} \log {\Delta_{sat} \over \Delta_k(T_{k+1})}
\label{eq:gotto}
\end{equation}
so that from (\ref{eq:gsei}), (\ref{eq:gsette}), 
(\ref{eq:gotto}) and using the fact that 
$e^{M_k \lambda_k}\Delta_k(T_{k+1})\sim \Delta_{sat}$ we obtain
\begin{equation}
M_k = {1 \over \lambda_k} \log {(\alpha-1)\lambda_{k} \over \epsilon}, \qquad
\alpha=2^{2/3}.
\label{eq:gnove}
\end{equation}

Figure \ref{fig:fig2} shows the result obtained
averaging over $1000$ initial perturbations.
Scattered markers represent the values of $\Delta_1(T_k)$
at the different saturation times, obtained according to the quasi-linear 
analysis. The agreement with the direct simulation of
(\ref{eq:guno}) is quite good.
The global behavior can be fitted by a power law, whose slope can be roughly
estimated by the following argument.
From (\ref{eq:gnove}), one has  that the periods $M_k$ scale, a
part logarithmic corrections, as $M_k \sim \lambda_k^{-1} \sim \alpha^{-k}$.
From (\ref{eq:gsette}) the growth of $\Delta_1$ during period
$M_k$ is estimated to be $\sim \epsilon^{k}$. Then, we expect 
$\Delta_1(t) \sim t^{\gamma}$ with $\gamma=- \log \epsilon / \log \alpha$.

We  have also performed a second  numerical experiment with a stronger
coupling $\epsilon=10^{-2}$ and quadratic  
coupling function $f(x)=x^2$. The initial error is $\delta_0=10^{-3}$, while all the 
other parameters in the model are unchanged.
We take  an average of $1000$ runs of $2000$ steps in order to
get a good statistics. The results are reported in figure \ref{fig:fig3},
 where one sees that the error on large scale $\Delta_1(t)$ has a complex
behaviour resulting from the combinations of several time-scales, which
are now not well separated. However, even in this case, it is possible to
determine an apparent power-law regime for intermediate times.

The system of coupled maps with different chaotic
characteristic time teaches us that it is possible
to have non trivial time evolutions of {\em non-infinitesimal} 
 perturbations at slowest
scale, and that these can be fitted by power laws,
although they are actually generated by saturation
processes. In this kind of systems, if we are interested in long time
predictions, the maximum Lyapunov exponent is of little or no importance and one
should consider all the time scales present in the system.

\section{Predictability in a Shell Model of turbulence}
\label{sec:Shellm}
In this section we discuss the problem of predictability in a model 
for the energy cascade in three dimensional fully developed 
turbulence \cite{Yamada,JPV91,ABCFPV94,LIMA}. The model, which is intended to mimic the 
Navier-Stokes equations, is defined in the Fourier space as follows.
The Fourier space is divided into $N$ shells labelled by the wavevector 
modulus $k_n = k_0\,2^n$, where $k_0$ is an arbitrary constant, each one 
containing all the wavevector $\bbox{k}$ with modulus 
$k_0\, 2^n < |\bbox{k}| \leq k_0\, 2^{n+1}$. The velocity difference over 
the length scale $\sim k_n^{-1}$ is given by a complex variable $u_n$ 
representing the Fourier components of the velocity field. The evolution 
is given by the set of $N$ ordinary differential equation
\begin{eqnarray}
 \frac{d}{dt} u_n &=& - \nu k_n^2 u_n + i g_n + f_n
\label{eq:sh1} \\
  g_n &=& k_n\,     u_{n+1}^*\, u_{n+2}^* -
         {1 \over 2}\, k_{n-1}\, u_{n-1}^*\, u_{n+1}^* -
         {1 \over 2}\, k_{n-2}\, u_{n-2}^*\, u_{n-1}^*
\label{eq:sh3}
\end{eqnarray}
where $f_n$ is an external forcing, acting on large scale, necessary to have a 
stationary state, and $\nu$ the viscosity.
The main differences with the Navier-Stokes equations are:
\begin{itemize}

\item{i)} the wave-vectors and the velocity fields $u_n$ are scalars;

\item{ii)} there are only nearest and next-nearest neighbour interactions
among shells.

\end{itemize}
The first point implies that it is not possible to have geometrical 
structures, all informations on phases being lost. The second point does 
not represent a strong limitation as long as the energy cascade in Fourier 
space is local, with exponentially decreasing interactions among shells. 
This is rather sensible for three-dimensional turbulence, but not for 
the two-dimensional case \cite{ABCFPV94}.

Despite the fact that the time evolution governed by (\ref{eq:sh1})-(\ref{eq:sh3}) 
spends long times around an unstable fixed point given by the Kolmogorov law 
$|u_n|\propto k_n^{-1/3}$ \cite{LIMA}, it 
exhibits chaotic behaviour on a strange attractor in 
the $2N$ dimensional phase space, with maximum Lyapunov exponent roughly 
proportional to $\nu^{-1/2}$. The velocity structure functions
 $\langle|u_n|^p\rangle \sim k_n^{-\zeta_p}$ have non-linear exponents $\zeta_p$
 very similar to those of real turbulence \cite{JPV91}.

Our study of predictability is based on the comparison of the temporal 
evolution of pairs of different realizations of the velocity field, 
say $u_n$ and $v_n$. 
Both fields evolve according to (\ref{eq:sh1})-(\ref{eq:sh3}) form
initial conditions such that:

\begin{itemize}
\item{i)}
the energy spectra of $u_n$ and $v_n$ at the initial time are equal;

\item{ii)}
$u_n$ and $v_n$ at initial time differ only on small scales, corresponding 
to wavenumber $k_n \geq k_{n_{\rm E}}$:
\begin{equation}
\label{eq:vn0}
 v_n = \left\{\begin{array}{ll}
                     u_n & \mbox{\rm for}\ n < n_{\rm E} \\
                     e^{i w_n}\, u_n  & \mbox{\rm for}\ n \geq n_{\rm E}
                    \end{array}\right.
\end{equation}
where $w_n$ is a random number uniformly distribited in the range $[0,\theta]$. 
\end{itemize}

By changing the value of $\theta$ we can modify the correlation between 
the two field. The extreme case $\theta= 2\pi$, i.e. completely 
uncorrelated variables for $n\geq n_{\rm E}$, corresponding to the
Lorenz choice discussed in section \ref{sec:Somer}.

Previous works \cite{CJPV93} investigated the growth of different small
perturbations in shell models. Here we want to study this issue for
a finite perturbation at initial time. 

In the numerical simulation we have taken as the reference field 
$u_n$ in (\ref{eq:vn0}) a solution of (\ref{eq:sh1})-(\ref{eq:sh3}) 
obtained from a long simulation starting from a random 
initial condition. The forcing $f_n$ was taken constant and equal to
\begin{equation}
\label{eq:forcing}
  f_n = (1+i)\,\gamma\,\delta_{n,4}
\end{equation}
with $\gamma= 0.005$.
The simulations were done mainly for two different systems: 
(A) $N=19$ and $\nu=10^{-6}$; (B) $N=27$ and $\nu=10^{-9}$. 
We used the leap-frog method of integration with time step
$10^{-3}$ in the case (A) and $2\times 10^{-5}$ in the case (B).
The average is over $10^3$ different realizations of the 
field $v_n$. All the figures will be for the case (A), the case
(B) being similar.

In order to understand whether intermittency has any effect on the
qualitative features, we also performed a detailed study in terms of
closure approximation, where, by construction, one considers averaged
quantities neglecting intermittency.  We thus developed the
standard eddy damped quasi-normal markovian  (EDQNM) approximation \cite{Lesieur}
 for the shell model (see appendices 
\ref{app:A} and \ref{app:B}).

Among the various quantities that can be computed, we focused on the 
error-spectrum at wavevector $k_n$
\begin{equation}
\label{eq:errsp}
 \Delta_n = \frac{1}{2}\,\langle (u_n - v_n)\,(u_n^*-v_n^*) \rangle
          = (E_n - {\rm Re}\,W_n)
\end{equation}
where $E_n = \langle u_n\,u_n^* \rangle =  \langle v_n\,v_n^* \rangle$ is the
energy of the two fields and $W_n = \langle u_n\,v_n^* \rangle$ is the
overlap energy at scale $n$. 

In Figs. \ref{fig:fig4} and \ref{fig:fig5} the error energy spectrum
$\Delta_n$ is showed at different times for $\theta=2\pi$ and $10^{-4}$, 
respectively.

In figure \ref{fig:fig4} we see that at the beginning we have a very
fast growth until $\Delta_n$ reaches the saturation (i.e.
$\Delta_n=E_n$) at large $n$. Then one has a sort of inverse cascade on
the error. This is also the main feature one observes in the case of
strong initial perturbation (figure \ref{fig:fig5}).

In order to compare the error growth in the shell model with the
behaviour observed in the toy model of section \ref{sec:Maps}, we also
compute the (global) growth of the moments of the difference field,
defined as
\begin{equation}
\label{eq:moments}
 \langle |\delta u|^q \rangle = \left\langle \left[ \sum_n\, 2 \Delta_n\right]^{q/2}
                           \right\rangle.
\end{equation}
and the large-scale error
\begin{equation}
\delta U_0 = \left[ \frac{\Delta_6 + \Delta_7 + \Delta_8}{3} \right]^{1/2}
\label{eq:lserror}
\end{equation}

Figures \ref{fig:fig6} and \ref{fig:fig7} show the growth of
the moments of $\langle |\delta u|^q\rangle^{1/q}$ respectively for 
$\theta=2\pi$ and
$\theta=10^{-4}$. For the first case we observe at intermediate times a
power law behaviour in agreement with the Lorenz prediction. The
intermittency produces anomalous scaling, i.e. $\langle |\delta u|^q
\rangle \sim t^{q \alpha(q)}$ where $\alpha(q)$ is a non increasing
function of $q$. Let us note that $\alpha(2)=1/2$, as in Lorenz
prediction where no intermittent effects are taken into account.
In the case of small initial perturbation (figure \ref{fig:fig7})
one can recognize an initial exponential growth until the error
saturates at small scales. Then we find a kind of power law behaviour. 

For a qualitative comparison with the results discussed for the coupled
maps, we show in figure \ref{fig:fig8} the growth of $\delta U_0$. We can
recognize an initial exponential growth driven by the smallest scale; in
later times all the time scales interplay leading to a non simple
function which can be misinterpreted as a power law.

The closure approximation for the shell model leads to a qualitaive 
similar behaviour. The basic idea of closure
is quite straightforward: one writes a Reynolds hierarchy for the
moments of the shell variables, and truncate the chain of equations at
the lowest sensible order. The full derivation is given below in
appendices \ref{app:A} and \ref{app:B}. The closure equation thus gives
information of the behaviour of $E_n$ and $W_n$ as functions of time.
The results are similar to those obtained by the direct integration (and
ensamble average) of the shell model, as is evident by the direct
comparison of figure \ref{fig:fig9} with figure \ref{fig:fig4} and
figure \ref{fig:fig10} with \ref{fig:fig5}. 
This is a clear evidence that the relevant mechanism, at least at a
qualitative level,  is due to the existence of many
characteristic times and not to intermittency effects.

\section{Discussion}
\label{sec:Discussion}
The predictability problem can be formulated in terms of the Lyapunov exponent, 
or the effective Lyapunov exponent, only for infinitesimal perturbations. 
On the other hand in systems with many different characteristic times and
finite perturbations the non-linear effects are very relevant. Direct 
consequence is that the growth of the perturbation is, a part at very
short times, not exponential but rather similar to a power law. Actually 
it is a sequence of exponentials with varying rates. This mechanism has
an important practical relevance: the predictability time for finite perturbations
can be much larger than the inverse of the Lyapunov exponent (\ref{eq:dx}). 

For example, in three dimensional turbulence for a ``physical'' perturbation 
localized at the initial time at the  Kolmogorov lenght $\eta$, 
$\delta\bbox{v}= O(v(\eta))$, the predictability time is $T \sim L/V$, 
independent of the Reynolds number. On the contrary, the predictability 
time for an
infinitesimal perturbation is proportional to the 
inverse of the maximum Lyapunov exponent $\lambda$, which grows
as a power of the Reynolds number. Hence the predictability time
decreases for increasing $Re$. 

These results give a strong indication that despite 
the presence of strong chaos, realistic situations have relatively long 
predictability times. 

In conclusion we have showed 
that the concepts of Lyapunov exponents, 
effective Lyapunov exponent, Kolmogorov entropy give insufficient informations
on the chaotic behaviour
of extended systems, and it is urgent the identification of new
indicators.

\section*{Acknowledgments}
This work was supported by the Swedish Natural Science Research Council
under contracts S-FO 1178-302 (E.A.) and by the INFN (Iniziativa
Specifica Meccanica Statistica FI3). E.A. thanks the Dipartimento di
Fisica, Universit\'a di Roma ``La Sapienza'' for hospitality and financial
support. G.B. thanks the ``Istituto di Cosmogeofisica del CNR'', Torino,
for hospitality.

\appendix
\section{}
\label{app:A}
Here we derive the equations for the energy of the field 
in the eddy damped quasi-normal markovian approximation (EDQNM)
for the shell model. For more details see e.g. \cite{Lesieur,Orszag}.

We are interested into the energy of the shell $n$ given by 
$E_n = \langle u_n\,u_n^*\rangle$. Differentiation with respect to time of $E_n$, 
and use of (\ref{eq:sh1})-(\ref{eq:sh3}), leads to the evolution equation for
$E_n$ for the shell model:
\begin{equation}
\label{eq:An}
  \frac{d}{dt} E_n = - 2\,\nu\,k_n^2\, E_n + i\langle g_n\,u_n^*\rangle
                                    - i\langle u_n\,g_n^*\rangle
                                    + 2\,\epsilon\,\delta_{n,4}.
\end{equation}
In order to have a constant energy input $\epsilon$ we assumed a forcing 
\begin{equation}
\label{eq:newforc}
  f_n = \frac{u_n}{|u_n|^2}\,\epsilon\,\delta_{n,4}
\end{equation}
in (\ref{eq:sh1}).
This equation is not closed since the third and fourth terms on r.h.s
involve third order correlation functions. For example
\begin{equation}
\label{eq:gust}
 \langle g_n\,u_n^*\rangle = 
                k_n     \langle u_n^*\,     u_{n+1}^*\, u_{n+2}^*\rangle -
   {1 \over 2}\,k_{n-1} \langle u_{n-1}^*\, u_{n}^*\,   u_{n+1}^*\rangle -
   {1 \over 2}\,k_{n-2} \langle u_{n-2}^*\, u_{n-1}^*\,u_{n}^*\rangle
\end{equation}
and similarly the other. The third order cumulants obey a differential 
equation involving fourth order correlation functions, e.g.
\begin{equation}
\label{eq:fourth}
\begin{array}{ll}
 \frac{\displaystyle d}{\displaystyle dt}
  \langle u_n\,u_{n+1}\,u_{n+2}\rangle = 
  -&\nu\,(k_n^2 + k_{n+1}^2 + k_{n+2}^2) \langle u_n\,u_{n+1}\,u_{n+2}\rangle \\
           &+i\langle g_n\,u_{n+1}\,u_{n+2}\rangle
            +i\langle u_n\,g_{n+1}\,u_{n+2}\rangle
            +i\langle u_n\,u_{n+1}\,g_{n+2}\rangle,
 \end{array}
\end{equation}
and so on. Thus in order to have a closed problem we have to truncate
the hierarchy. This is done assuming a quasi-normal probability distribution
for the field, and factorizing the fourth order correlation functions as
\begin{equation}
\label{eq:QN}
  \langle x_1\,x_2\,x_3\,x_4\rangle = 
        \langle x_1\,x_2\rangle\,\langle x_3\,x_4\rangle +
        \langle x_1\,x_3\rangle\,\langle x_2\,x_4\rangle +
        \langle x_1\,x_4\rangle\,\langle x_3\,x_2\rangle.
\end{equation}
By means of (\ref{eq:QN}), and the assumption of isotropic turbulence
\begin{equation}
\label{eq:isotrop}
 \langle u_n\,u_m^* \rangle = \delta_{nm}\, E_n, \qquad
 \langle u_n\,u_m \rangle = 0
\end{equation}
it is easy to evaluate the terms involving the fourth order correlation
functions in the equations for the third order ones, and close the hierarchy.
For example for Eq. (\ref{eq:fourth}) we have,
\begin{equation}
\label{eq:f1}
\begin{array}{ll}
 \langle g_n\,u_{n+1}\,u_{n+2}\rangle =& k_n\, E_{n+1}\,E_{n+2} \\
 \langle u_n\,g_{n+1}\,u_{n+2}\rangle =& -{1 \over 2}\,k_n\, E_{n} \,E_{n+2} \\
 \langle u_n\,u_{n+1}\,g_{n+2}\rangle =& -{1 \over 2}\,k_n\, E_{n} \,E_{n+1}. \\
 \end{array}
\end{equation}
Inserting (\ref{eq:f1}) into (\ref{eq:fourth}) we obtain
%\begin{equation}
%\label{eq:thirdQN}
%  \begin{array}{ll}
%  \left[ \frac{\displaystyle d}{\displaystyle dt} + 
%         \nu (k_n^2 + k_{n+1}^2 + k_{n+2}^2 ) \right]
%       &\langle u_n\,u_{n+1}\,u_{n+2}\rangle = \\
%                            & i a\,k_n\, E_{n+1}\,E_{n+2} +
%                              i b\,k_n\, E_{n}  \,E_{n+2} +
%                              i c\,k_n\, E_{n}  \,E_{n+1}
%  \end{array}
%\end{equation}
\begin{equation}
\label{eq:thirdQN}
  \left[ \frac{\displaystyle d}{\displaystyle dt} + 
         \nu (k_n^2 + k_{n+1}^2 + k_{n+2}^2 ) \right]
       \langle u_n\,u_{n+1}\,u_{n+2}\rangle = i S(n,t)
\end{equation}
with
\begin{equation}
\label{eq:S}
  S(n,t) = k_n\, E_{n+1}\,E_{n+2} -
                   {1 \over 2}\,k_n\, E_{n}  \,E_{n+2} -
                   {1 \over 2}\,k_n\, E_{n}  \,E_{n+1}.
\end{equation}
Similar equations hold for the third order cumulants entering into 
Eq. (\ref{eq:gust}).

The quasi-normal approximation (\ref{eq:thirdQN}) it is known to give
unphysical results \cite{Orszag}. In particular it does not produce 
positive definite energy spectra.
The problem has been cured by introducing the so-called ``eddy damped''
approximation. One replaces (\ref{eq:thirdQN}) with
%\begin{equation}
%\label{eq:thirdEDQN}
% \begin{array}{rcl}
%  \Bigl[ \frac{\displaystyle d}{\displaystyle dt} + 
%          \nu (k_n^2 + k_{n+1}^2 + k_{n+2}^2 ) &+&
%          \mu_n + \mu_{n+1} + \mu_{n+2} \Bigr]
%       \langle u_n\,u_{n+1}\,u_{n+2}\rangle = \\
%                            &&  i a\,k_n\, E_{n+1}\,E_{n+2} +
%                              i b\,k_n\, E_{n}  \,E_{n+2} +
%                              i c\,k_n\, E_{n}  \,E_{n+1}
% \end{array}                              
%\end{equation}
\begin{equation}
\label{eq:thirdEDQN}
 \begin{array}{rcl}
  \Bigl[ \frac{\displaystyle d}{\displaystyle dt} +
          \nu (k_n^2 &+& k_{n+1}^2 + k_{n+2}^2 ) \\ &+&
          \mu_n + \mu_{n+1} + \mu_{n+2} \Bigr]
       \langle u_n\,u_{n+1}\,u_{n+2}\rangle = i S(n,t)
  \end{array}
\end{equation}
where 
\begin{equation}
\label{eq:mu}
 \mu_n \equiv \mu(k_n,E_n) = \alpha\,k_n\,E_n^{1/2}.
\end{equation}
Dimensionally $k_n E_n^{1/2}$ is an inverse time, the turn-over time at
shell $n$. We have one free parameter, the dimensionless constant
$\alpha$. It should be adjusted such that the spectrum is as similar as
possible to the spectrum obtained in simulations of the full equation.
We will return to this point later.

Equation (\ref{eq:thirdEDQN}) can be easily integrate, and the solution
reads
\begin{equation}
\label{eq:trdEDQN}
  \langle u_n\,u_{n+1}\,u_{n+2}\rangle = i\int_{0}^{t}\, dt'\,
   e^{-[ \nu (k_n^2 + k_{n+1}^2 + k_{n+2}^2 ) +
                      \mu_n + \mu_{n+1} + \mu_{n+2} ]\,(t-t')
     }\, S(n,t').
\end{equation}
It has been proved that if we assume that $S(n,t)$ does not vary 
significantly in the range where the exponential in (\ref{eq:trdEDQN}) is
substantially different from zero, then this is 
a sufficient condition for the positiveness of 
the energy spectra \cite{Orszag}. This assumption is called Markovianization. 
Therefore
the third order correlation function at time $t$ in the EDQNM 
approximation reads
\begin{equation}
\label{eq:trdEDQNM}
  \langle u_n\,u_{n+1}\,u_{n+2}\rangle = i\theta(n,t)\,
                                          S(n,t)
\end{equation}
where
\begin{eqnarray}
  \theta(n,t)& = & \int_{0}^{t}\, dt'\,
               e^{-[ \nu (k_n^2 + k_{n+1}^2 + k_{n+2}^2 ) +
                      \mu_n + \mu_{n+1} + \mu_{n+2} ]\,(t-t')}
       \nonumber \\
 	& = & \frac{\displaystyle 1 - e^{-[ \nu (k_n^2 + k_{n+1}^2 
 	                                        + k_{n+2}^2 ) +
                      \mu_n + \mu_{n+1} + \mu_{n+2} ]\,t}
                }
                {\displaystyle 
                     \nu (k_n^2 + k_{n+1}^2 + k_{n+2}^2 ) +
                      \mu_n + \mu_{n+1} + \mu_{n+2}
                  }.
 	\label{eq:theta}
\end{eqnarray}                                         
Similar equations hold for the other third order correlation functions.

We can now go back to (\ref{eq:An}) and write down the equation for the
energy $E_n$ in the EDQNM approximation
\begin{equation}
\label{eq:EnEDQNM}
 \begin{array}{rl}
 \left(\frac{d}{dt} + 2\nu k_n^2\right)\, E_n = 2&\Bigl[
      k_n^2\,    \theta(n,t)\, S(n,t) \\ &\,-
      {1 \over 2}\,k_{n-1}^2\,\theta(n-1,t)\, S(n-1,t) \\ &\,-
      {1 \over 2}\,k_{n-2}^2\,\theta(n-2,t)\, S(n-2,t) \Bigr] 
      + 2\,\epsilon\,\delta_{n,4}.
 \end{array}
\end{equation}

The quasi-normal ansatz implies the absence
of intermittency corrections. This is an essential limitation of all
closure theories. The energy spectrum of the shell model in the
EDQNM approximation must therefore obey $E_n \simeq C(\alpha)
\epsilon^{2/3} k_n^{-2/3}$ in the inertial range. The undetermined
function $C(\alpha)$ is the Kolmogorov constant.

On the other hand it has become clear in several independent
investigations that intermittency corrections exist in shell models. The
energy spectrum is therefore in reality more closely described by $E_n
\sim F(\epsilon)k_n^{-\zeta_2}$, where the exponent $\zeta_2$ has been
estimated to be $0.70$ \cite{JPV91}. The function
$F$ that gives the prefactor to the power law in the inertial range
should not depend on viscosity, but depends on the forcing
through $\epsilon$, the mean dissipation of energy per unit time, or,
equivalently, the mean energy input into the system from the force. In a
really large inertial range the two power-laws are not good
approximations to one other. The best that can be done is to demand that
the spectra agree as closely as possible at the upper end of the
inertial range. The disagreement at the lower end will then be
approximately $(k_L/k_0)^{\zeta_2-2/3}$. This is not a very large
discrepancy. Assuming an inertial range of twenty shells, that is a
scale separation of $10^6$, and $\zeta_2=0.70$, the mismatch is only a
factor $1.5$. In practice a number of $\alpha$'s have been tried and the
two spectra compared until a reasonable agreement is achieved.
For $\alpha={\bf ???}$ we obtain $C(\alpha)= 1.5$ which is the value 
observed both in simulations of the shell model and in experiments
\cite{Orszag}.

%%%%%%%%%%%%%%%%%%%%%%%%%%%%%%%%%%%%%%%%%%%%%%%%%%%%%%
\section{}
\label{app:B}
Here we derive the equations for the energy of the field difference
in the eddy damped quasi-normal markovian approximation (EDQNM)
for the shell model. The procedure is similar to that described in 
App. \ref{app:A} for the energy, so we only report the main equations.

We consider to independent realizations of the field, $u_n$ and $v_n$, with 
the same energy spectrum $E_n$,
both evolving according to (\ref{eq:sh1}) and (\ref{eq:sh3}). For 
simplicity of notation, the equation of motion of the field $v_n$ is rewritten 
as
\begin{equation}
\label{eq:vshellm}
 \begin{array}{rl}
 \frac{d}{dt} v_n &= - \nu\,k_n^2 + i h_n + \tilde{f}_n \\
  h_n &=               k_n\,     v_{n+1}^*\, v_{n+2}^* -
         \frac{1}{2}\, k_{n-1}\, v_{n-1}^*\, v_{n+1}^* -
         \frac{1}{2}\, k_{n-2}\, v_{n-2}^*\, v_{n-1}^* \\
  \tilde{f}_n &= \frac{\displaystyle v_n}{\displaystyle |v_n|^2}\,\epsilon\,\delta_{n,4}
 \end{array}
\end{equation} 

We are interested into the energy of the field difference at the shell 
$n$
\begin{equation}
\label{eq:Dn}
  \Delta_n = \frac{1}{2}\,\langle (u_n - v_n)\,(u_n^* - v_n^*) \rangle
           = (E_n - {\rm Re}\, W_n)
\end{equation}
where $W_n=\langle u_n\,v_n^* \rangle$. The evolution equation of $W_n$ 
is easily derived by differentiation with respect to time and reads,
\begin{equation}
\label{eq:Wn}
  \frac{d}{dt} W_n = - 2\,\nu\,k_n^2\, W_n + i\langle g_n\,v_n^*\rangle
                                    - i\langle u_n\,h_n^*\rangle
                                    + 2\,\epsilon\,\delta_{n,4}
\end{equation}
As the case of $E_n$ this equation is not closed since it involves third 
order cumulants of $u_n$ and $v_n$.
For example
\begin{equation}
\label{eq:gvst}
 \langle g_n\,v_n^*\rangle = 
                        k_n     \langle v_n^*\,     u_{n+1}^*\, u_{n+2}^*\rangle -
           \frac{1}{2}\,k_{n-1} \langle u_{n-1}^*\, v_{n}^*\,   u_{n+1}^*\rangle -
           \frac{1}{2}\,k_{n-2} \langle u_{n-2}^*\, u_{n-1}^*\,v_{n}^*\rangle
\end{equation}
and similarly the other. As done for the energy $E_n$, the hierarchy is 
closed the fourth order correlation functions. 
With a calculation similar to that of App. \ref{app:A}, one obtains
\begin{equation}
\label{eq:vthirdQN}
  \begin{array}{rl}
  \left[ \frac{\displaystyle d}{\displaystyle dt} + 
         \nu (k_n^2 + k_{n+1}^2 + k_{n+2}^2 ) \right]
       &\langle v_n^*\,u_{n+1}^*\,u_{n+2}^*\rangle = \\
            & -i k_n\, W_{n+1}^*\,W_{n+2}^* 
              +i {1 \over 2}\,k_n\, W_{n}  \,E_{n+2} 
              +i {1 \over 2}\,k_n\, W_{n}  \,E_{n+1}
  \end{array}
\end{equation}
and similar equations for the others. 

We then perform the eddy damped and markovian approximation, i.e.
add a damping term $(\mu_n+\mu_{n+1}+\mu_{n+2}) W_n$ to the l.h.s, where
$\mu_n$ is given by (\ref{eq:mu}), and we integrate the resulting equation in
the markovian approximation -- see App. \ref{app:A} for more details.
From (\ref{eq:vthirdQN}), it follows 
\begin{equation}
 \label{eq:tuvEDQNM}
  \begin{array}{rl}
  \langle v_n^*\,u_{n+1}^*\,u_{n+2}^*\rangle = -i \theta(n,t)\,k_n\,
          &[k_n\, W_{n+1}^*\,W_{n+2}^* \\ &\,-
             {1 \over 2}\,k_n\, W_{n}  \,E_{n+2} -
             {1 \over 2}\,k_n\, W_{n}  \,E_{n+1} ]
   \end{array}
\end{equation}     
with $\theta(n,t)$ given by (\ref{eq:theta}).
The other cumulants leads to similar equations.
Collecting all the terms one finally obtains the equation for $W_n$ in the 
EDQNM approximation
\begin{equation}
\label{eq:WnEDQNM}
 \begin{array}{rl}
 \left(\frac{d}{dt} + 2\nu k_n^2\right)\, W_n = 2&\Bigl[
 k_n^2\,    \theta(n,t)\, 
 (W_{n+1}^*\,W_{n+2}^* - {1 \over 2}\,W_n\,E_{n+2} - {1 \over 2}\,W_n\,E_{n+1})
 \\ &\,-
 {1 \over 2}\,k_{n-1}^2\,\theta(n-1,t)\, 
 (W_{n}\,E_{n+1} - {1 \over 2}\,W_{n-1}^*\,W_{n+1}^* - {1 \over 2}\,E_{n-1}\,W_n)
 \\ &\,-
 {1 \over 2}\,k_{n-2}^2\,\theta(n-2,t)\, 
 (E_{n-1}\,W_n - {1 \over 2}\,E_{n-2}\,W_n - {1 \over 2}\,W_{n-2}^*\,W_{n-1}^*)
 \Bigr] \\ &\,+
 2\,\epsilon\,\delta_{n,4}.
 \end{array}
\end{equation}

%%%%%%%%%%%%%%%%%%%%%%%%%%%%%%%%%%%%%%%%%%%%%%%%%%%%%%%%%%%%%%

\begin{figure}
\caption{Error growth $\Delta_1$ as a function of $t$ for $\epsilon=10^{-4}$. 
         The straight segments correspond to  exponential growth driven by 
         different dominant scales.
        }
\label{fig:fig1}
\end{figure}

\begin{figure}
\caption{$\langle\Delta_1\rangle$ as a function of $t$ for $\epsilon=10^{-4}$.
         The average is taken over $1000$ initial configurations.
         The symbols (dyamonds) are the estimated growth of the largest scale 
         $\Delta_1$ during different periods $M_k$ according to the quasi-linear 
         approximation. Dashed line: power law expected by the simple argument 
         described in the text.
        }
\label{fig:fig2}
\end{figure}

\begin{figure}
\caption{Error growth $\Delta_1$ for a non-linearly coupled system with
         $\epsilon=10^{-2}$. 
        }
\label{fig:fig3}
\end{figure}

\begin{figure}
\caption{Error spectrum $\Delta_n$ as a function of $n$ at different time for case
         (A) and $\theta=2 \pi$. The lines are taken at times:$0$, $1$, $2$, $3$,
         $4$, $5$, $10$, $15$, $20$, $25$ and $30$ seconds after the perturbation 
         has been done.
         The dashed line is the spectrum $E_n$.
        }
\label{fig:fig4}
\end{figure}

\begin{figure}
\caption{Error spectrum $\Delta_n$ as a function of $n$ at different time for 
         case (A) and $\theta=10^{-4}$. The lines are taken at times:$0$, $5$, $10$, 
         $15$, $20$, $25$, $30$, $35$, $40$, $45$ and $50$ seconds after the perturbation 
         has been done.
         The dashed line is the spectrum $E_n$.
        }
\label{fig:fig5}
\end{figure}

\begin{figure}
\caption{Moments of the difference field $\langle |\delta u|^{q} \rangle^{1/q}$
         ($q=1,2,3,4$)
         as a function of $t$ for case (A) and $\theta=2 \pi$.
        }
\label{fig:fig6}
\end{figure}

\begin{figure}
\caption{Moments of the difference field $\langle |\delta u|^{q} \rangle^{1/q}$ 
         ($q=1,2,3,4$)
         as a function of $t$ for case (A) and $\theta=10^{-4}$.
        }
\label{fig:fig7}
\end{figure}

\begin{figure}
\caption{$\delta U_0$ as a function of $t$ for the case (A).
        }
\label{fig:fig8}
\end{figure}

\begin{figure}
\caption{Error spectrum $\Delta_n$ as a function of $n$ at different time for 
         the EDQNM approximation and $\theta=10^{-4}$.
         The lines are taken at time interval $0.02$ seconds from the 
         the perturbation, first line.
         The dashed line is the spectrum $E_n$.
        }
\label{fig:fig9}
\end{figure}

\begin{figure}
\caption{$\langle |\delta u|^2 \rangle$ as a function of $n$ at different time for 
         the EDQNM approximation and $\theta=10^{-4}$. 
        }
\label{fig:fig10}
\end{figure}

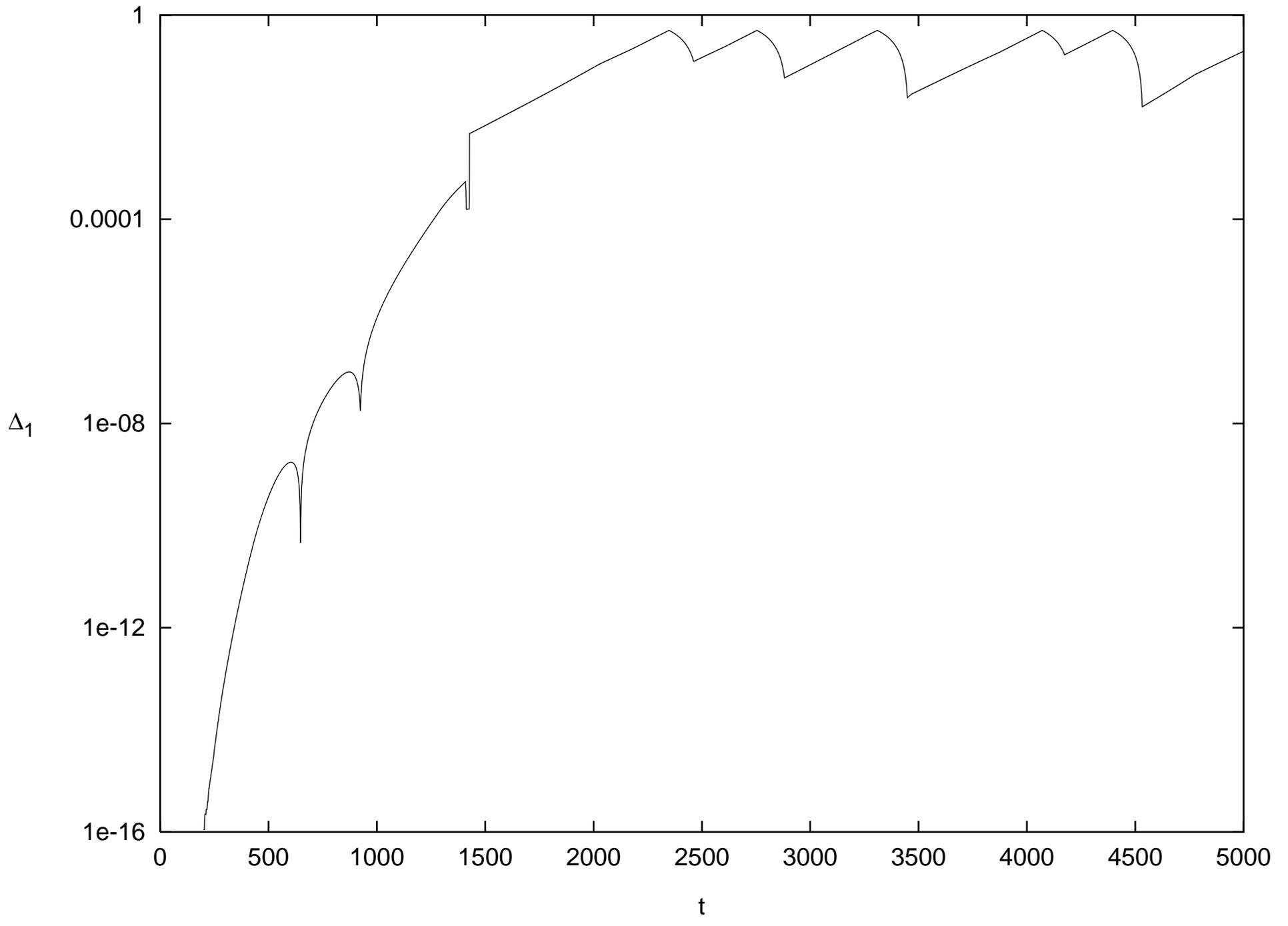
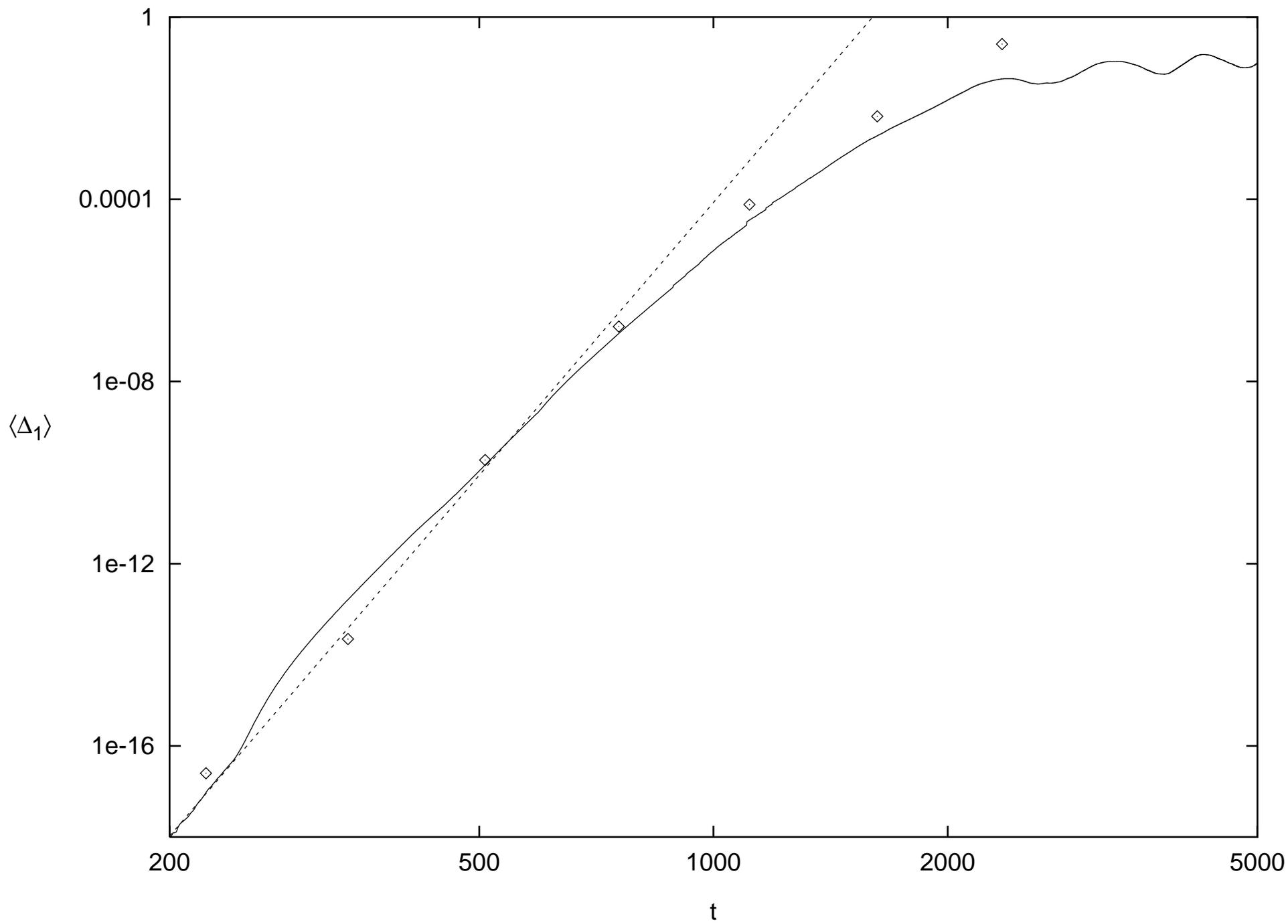
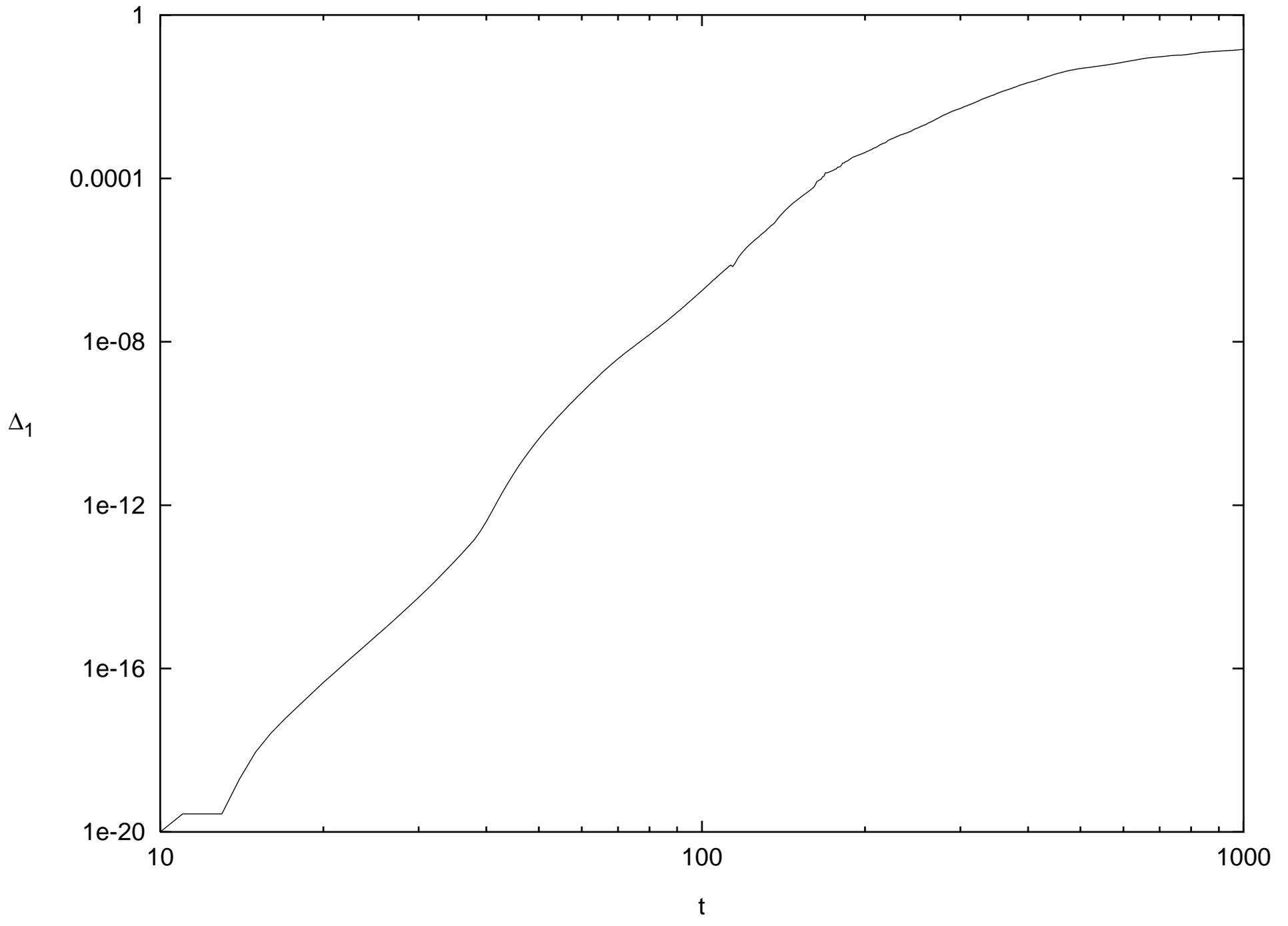
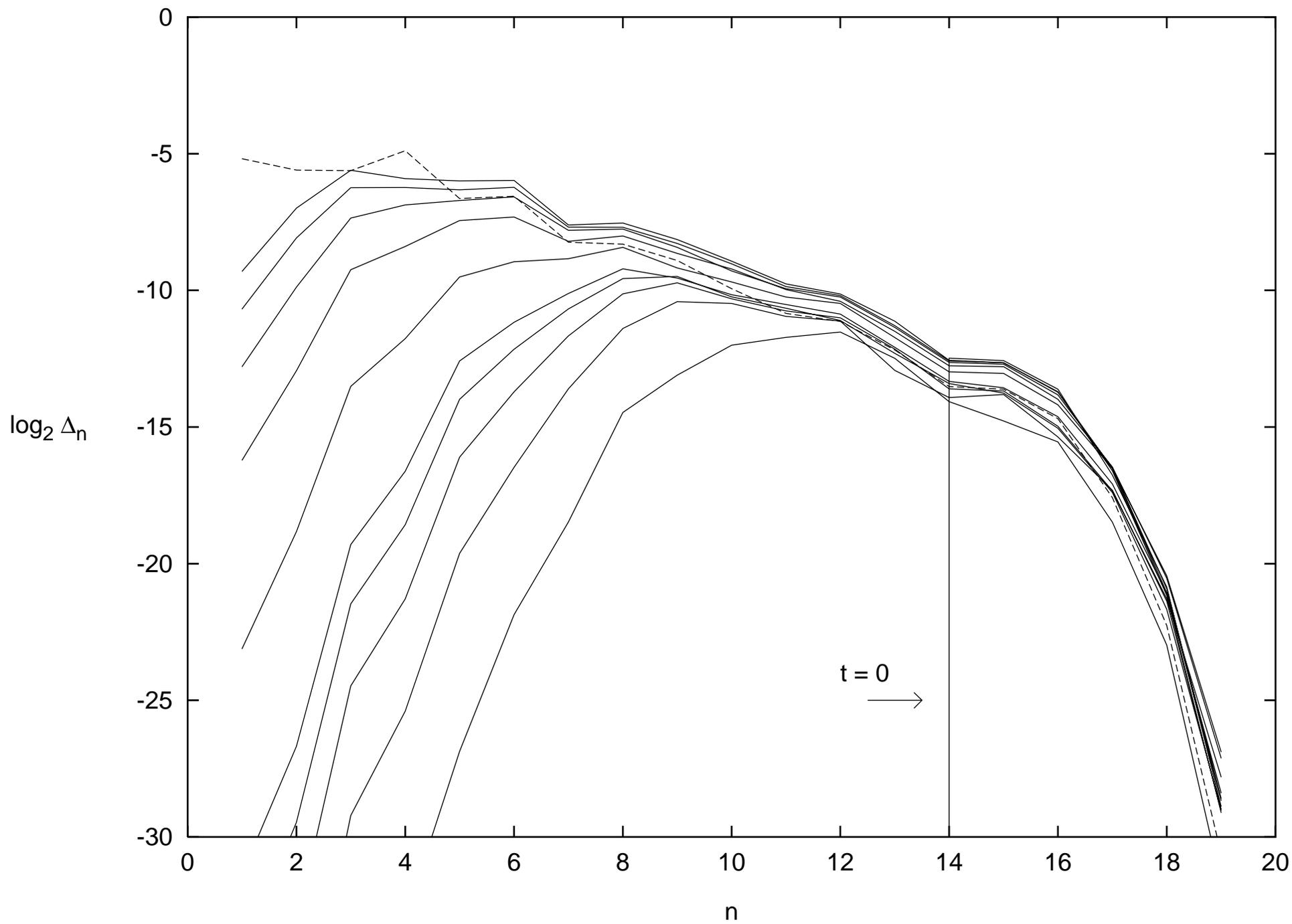
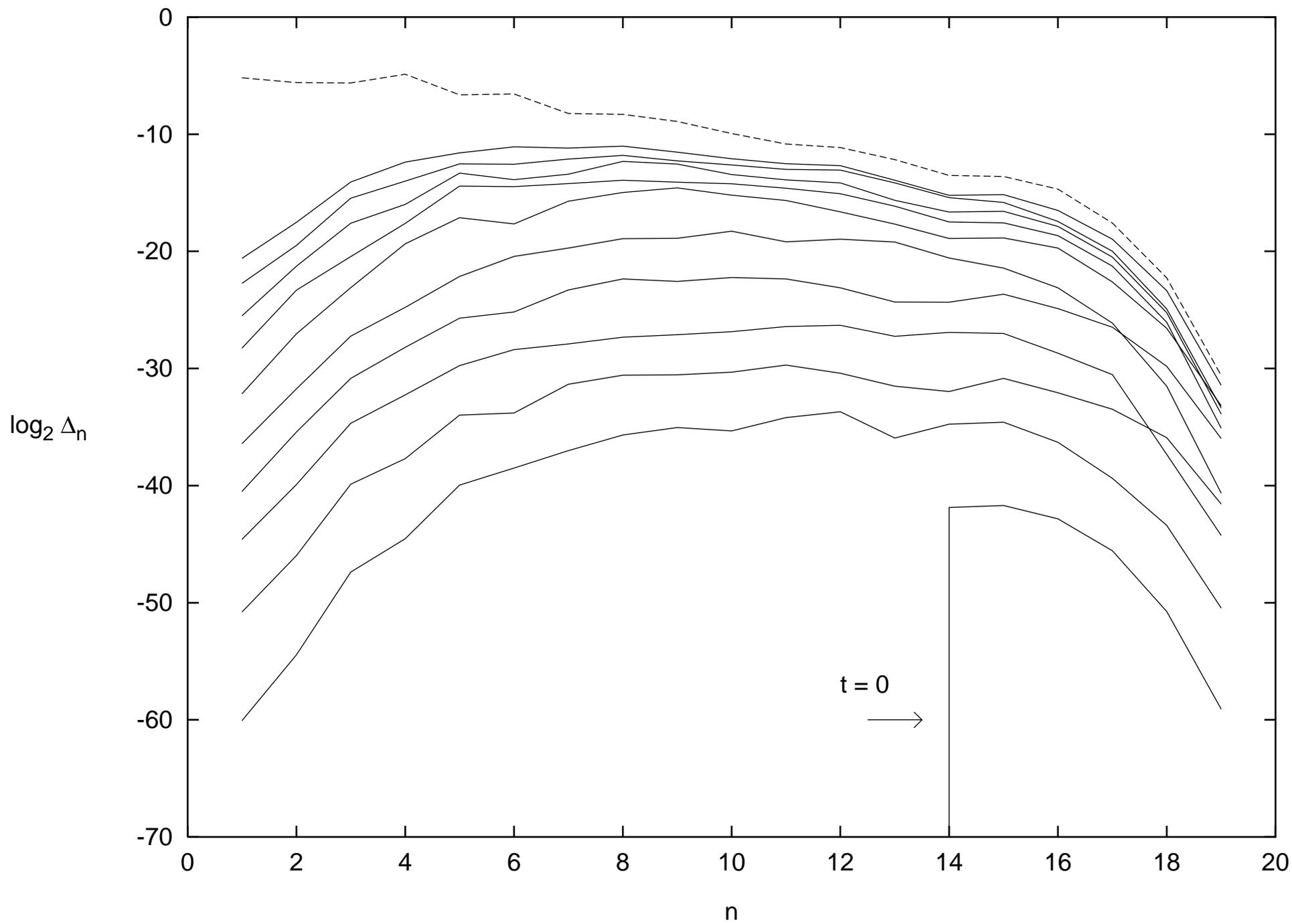
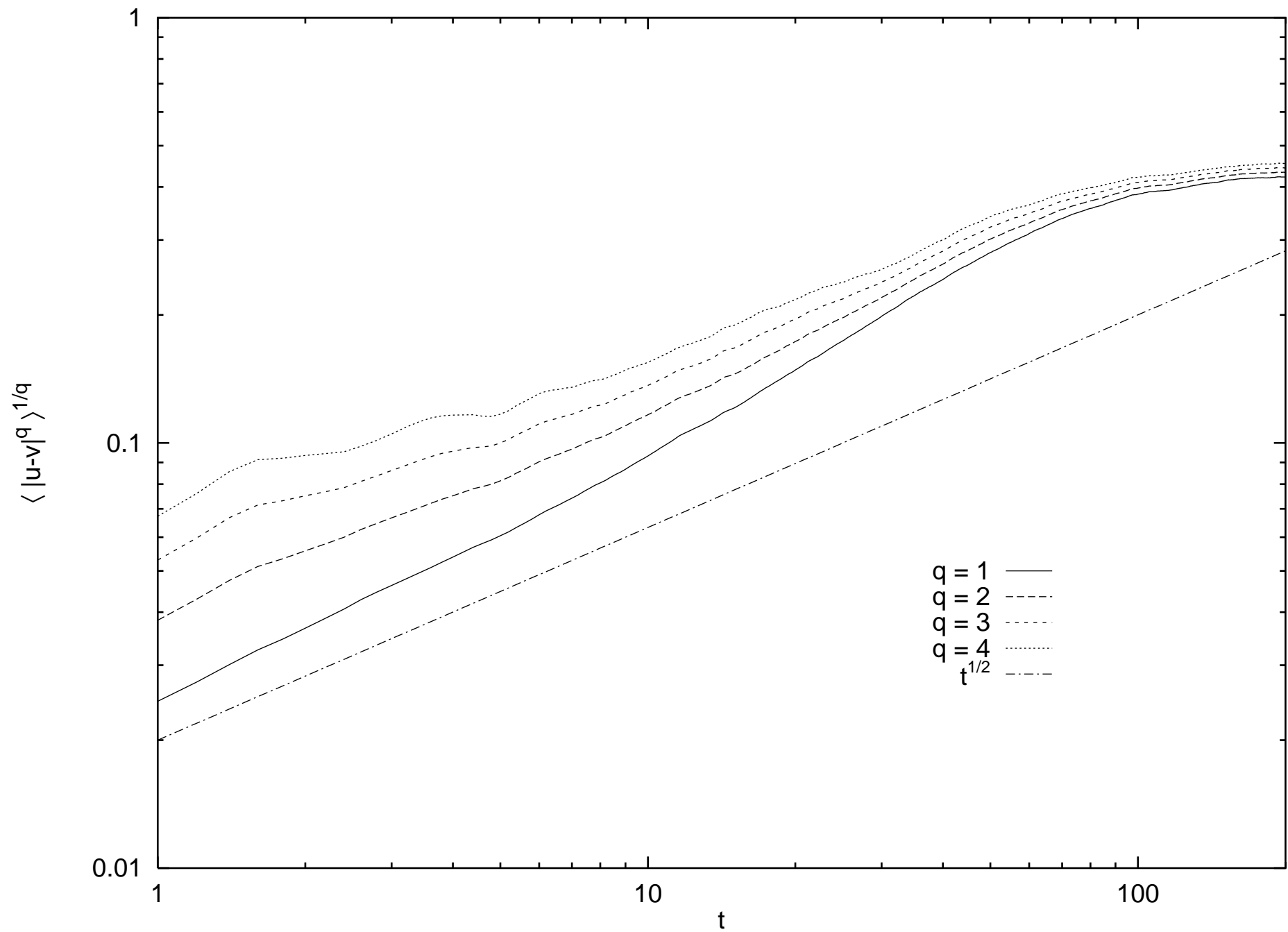
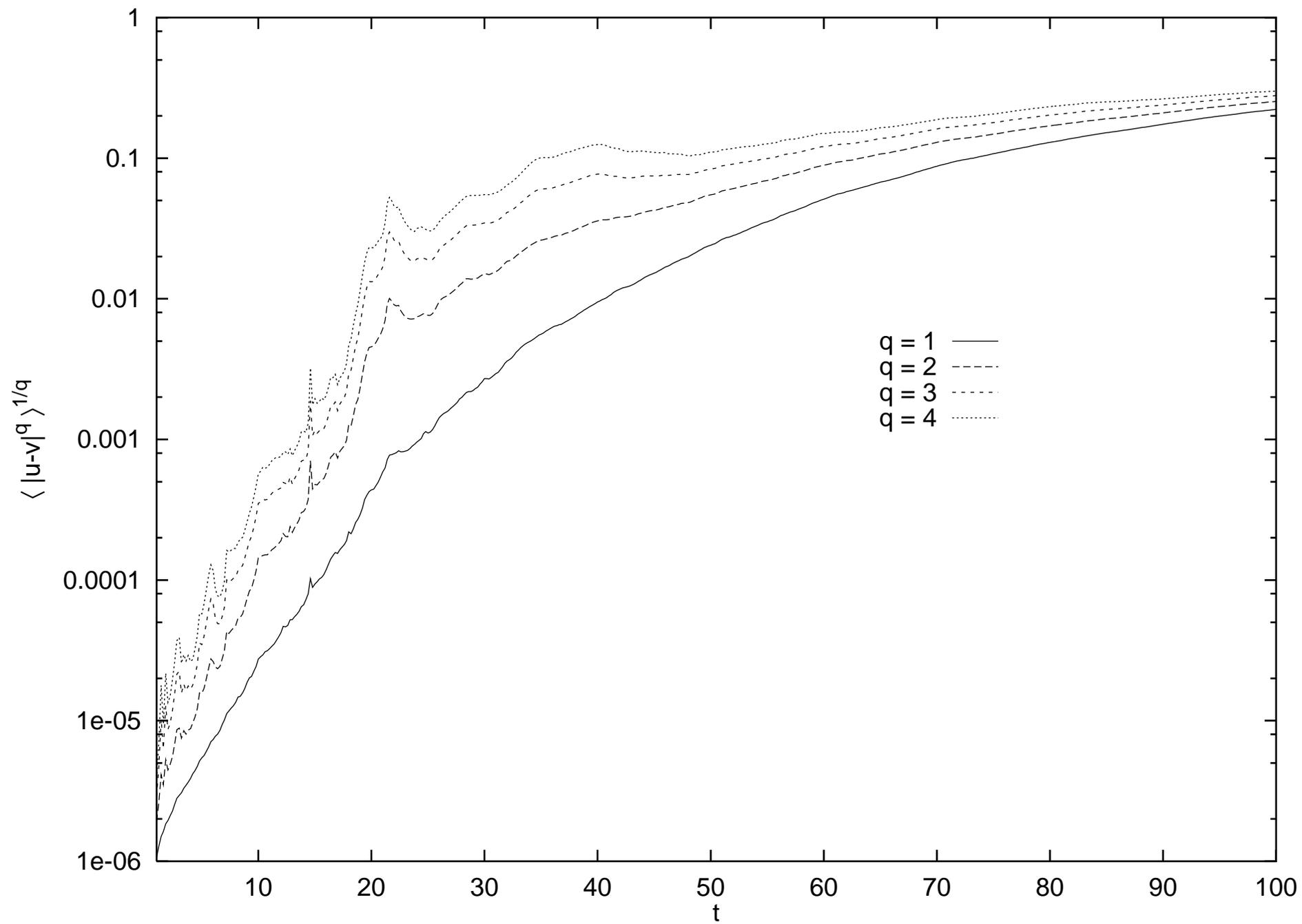
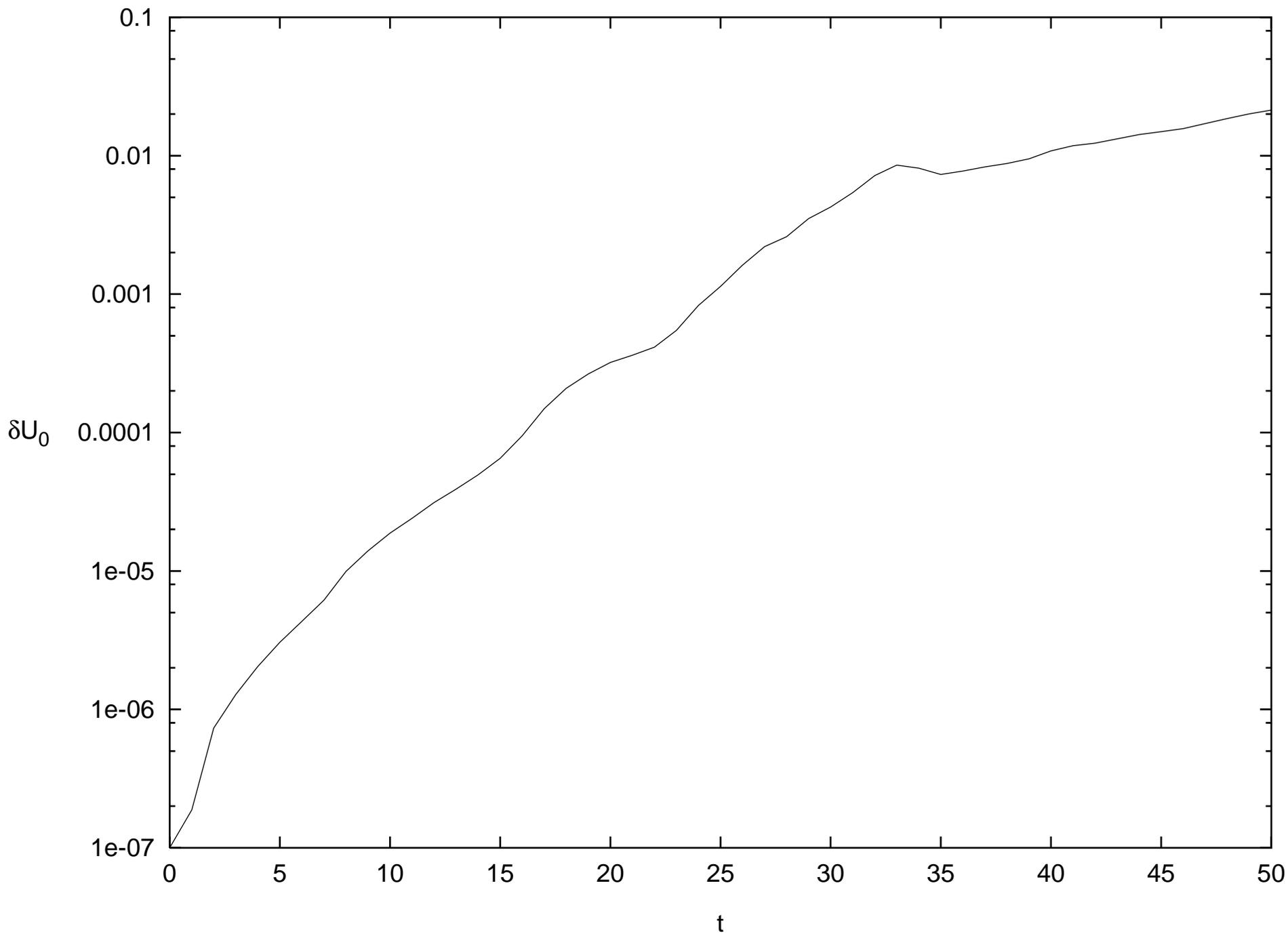
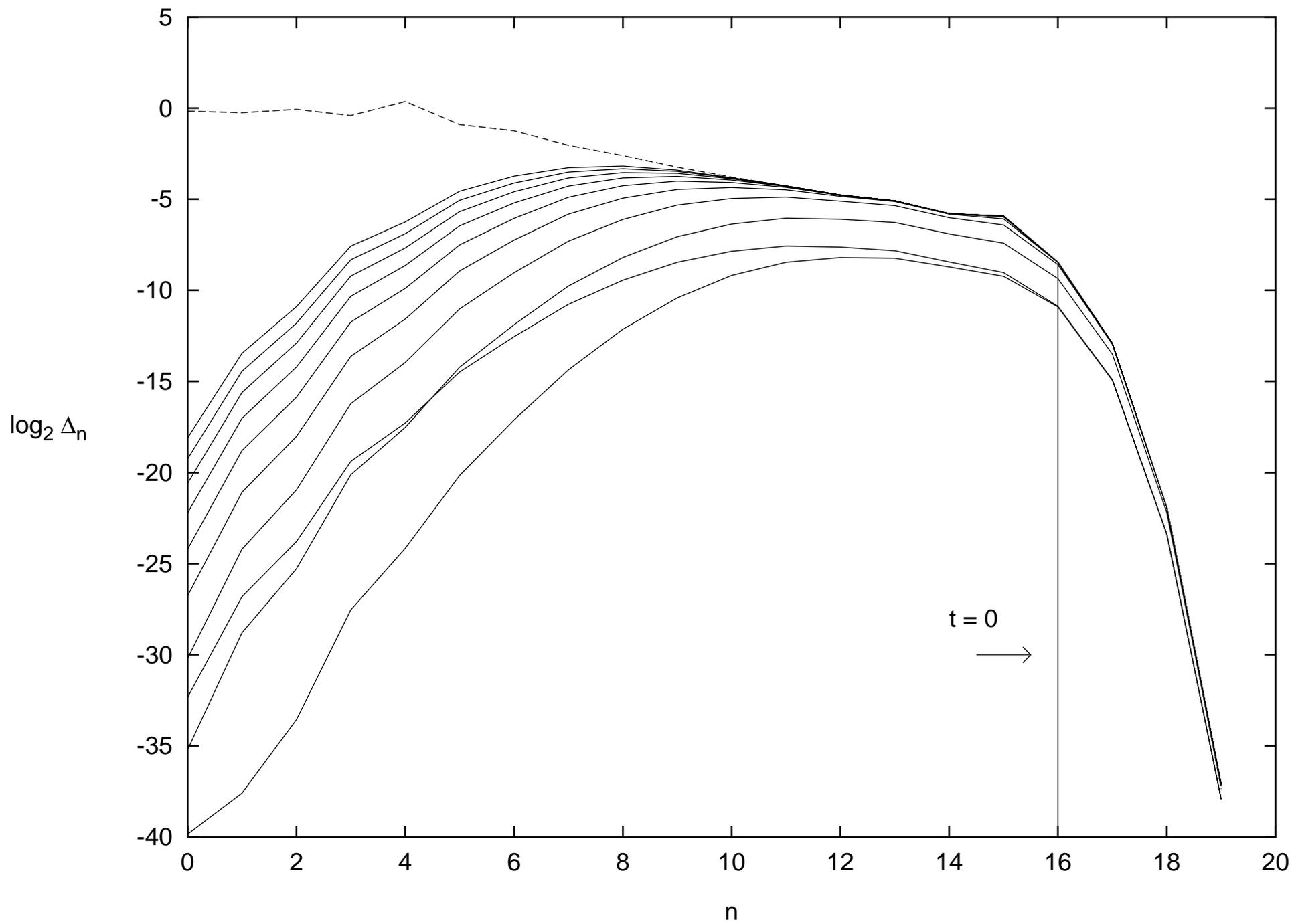
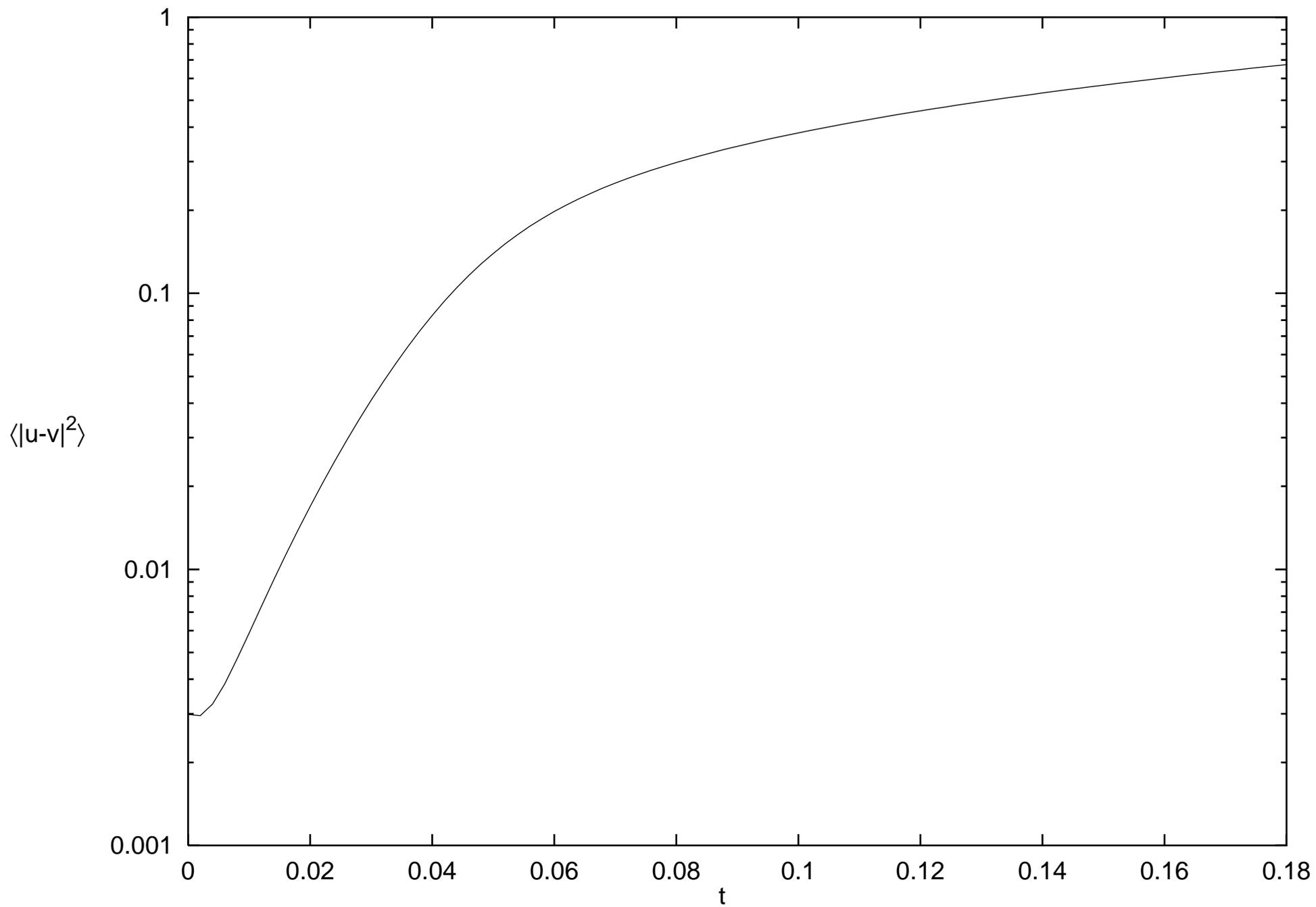
\end{document}